\documentclass{iaus}
\usepackage{graphicx}

\newcommand{\hMsun}{h^{-1} M_{\odot}}
\newcommand{\Msun}{M_{\odot}}

\title[Mergers and Disk Survival in $\Lambda$CDM] 
{Mergers  and Disk Survival in $\Lambda$CDM}

\author[Bullock, Stewart, and Purcell]   
{James. S. Bullock, Kyle R. Stewart, Chris W. Purcell}

\affiliation{Center for Cosmology, Department of Physics and Astronomy, The University of California, Irvine, CA 92697 USA}

\pubyear{2008}
\volume{254}  
\pagerange{1--4}
\jname{The Galaxy Disk in Cosmological Context}
\editors{J. Andersen, J. Band-Hawthorn, \& B. Nordstr{\"o}m, eds.}
\begin{document}

\maketitle

\begin{abstract}
Disk galaxies are common in our universe and this is a source of concern for  hierarchical formation models like $\Lambda$CDM.
Here we investigate this issue as motivated by  raw merger statistics derived for galaxy-size dark matter
halos from $\Lambda$CDM simulations.  Our analysis  shows that a majority ($\sim 70\%$) of  galaxy halos with
$M_0 = 10^{12} M_\odot$ at $z=0$ should have accreted  at least one object with mass $m  >   10^{11} \Msun \simeq 3 \, M_{\rm disk}$ over the last
$10$ Gyr.    Mergers involving larger objects $m \gtrsim 3 \times 10^{11} \Msun$ should have been very rare
for Milky-Way size halos today, and this pinpoints  $m/M \sim 0.1$ mass-ratio mergers
as the most worrying ones for the survival of thin galactic disks.
  Motivated by these results, we
use use high-resolution, dissipationless
$N$-body simulations to study the response of  stellar Milky-Way type disks
to these common mergers and show that thin disks
do not survive the bombardment.   The remnant galaxies
 are roughly three times as thick and twice
as kinematically hot as the observed thin disk of the Milky Way. 
Finally, we evaluate the suggestion that disks may be preserved
if the mergers involve gas-rich progenitors.  Using empirical measures to assign stellar masses and
gas masses to dark matter halos as a function of redshift, we show that the vast majority of
large mergers experienced by $10^{12} \Msun$ halos should be gas-rich ($f_{gas} > 0.5$), 
suggesting that this is a potentially viable solution to the disk formation conundrum.  
Moreover, gas-rich mergers should become increasingly rare in more massive halos $> 10^{12.5} \Msun$,
and this suggest that merger gas fractions may play an important role in establishing
morphological trends with galaxy luminosity.

\end{abstract}

\firstsection 
\section{Introduction}

Roughly 70\% of Milky-Way size dark matter halos  are believed to
host late-type, disk-dominated galaxies 
(Weinmann et al. 2006, van den Bosch et al. 2007, Ilbert et al. 2006, Choi et al. 2007).
Conventional wisdom dictates that disk galaxies result from fairly
quiescent formation histories, and this has raised concerns about disk formation 
within hierarchical Cold Dark Matter-based cosmologies (Toth \& Ostriker 1992; Wyse 2001; Kormendy et al. 2005).
Recent evidence for the existence of a sizeable population of cold, rotationally supported
disk galaxies at $z \sim 1.6$  ($\sigma/V \sim 0.2$; Wright et al. 2008) is particularly striking,
given that the fraction of galaxies with recent mergers is expected to be significantly higher
at that time (Stewart et al. 2008b).

Unfortunately, a real evaluation of the
 severity of the problem is limited by both theoretical and observational concerns.
Theoretically, the process of disk galaxy formation
remains very poorly understood in $\Lambda$CDM.   Though the first-order models
envisioned by Mestel (1963), Fall \& Efstathiou (1980), Mo, Mao \& White (1998) and others 
provide useful theoretical guides, the
formation of disks via a quiescent acquisition of mass is likely not the only channel.     
Over the last several years,
cosmological hydrodynamic simulations have begun to produce galaxies that resemble realistic
disk-dominated systems, and in most cases, early mergers have played a role in the disk's  formation
(Abadi et al. 2003; Brook et al. 2004; Robertson et al. 2005; Governato et al. 2007).  In particular,
the early disks in the simulations of Brook et al. (2004) and Robertson et al. (2005) originated in
gas-rich mergers.  Robertson et al. (2006) used a suite of focused simulations to show that
mergers with gas-fractions larger than $\sim 50\%$ tend to result in disk-dominated remnants  and
Hopkins
et al. (2008) used a larger sample of merger simulations to reach a similar conclusion.  
Two cautionary notes are in order.  First, these results are all subject to the uncertain assumptions
associated with modeling `subgrid' physics in the simulations
(ISM pressure, star formation, feedback, etc.).  Second,  the merger-remnant disks in these simulations
 tend to be hotter and thicker than the {\em thin} disk of the Milky Way
(Brook et al. 2004).  Robertson \& Bullock (2008) showed that gas-rich merger remnants
are a much closer match to the high-dispersion, rapidly rotating disk galaxies 
observed by integral field spectroscopy at $z\sim2$ (F{\"o}rster Schreiber et al. 2006; Genzel et al. 2006).  
Gas-rich mergers as an explanation for these high-redshfit disks is further motivated
by the expectation that gas fractions should be higher at early times (e.g. Erb et al. 2006).

Observationally, the best quantified thin disk is that of the Milky Way.
The thin disk of the
Milky Way  has a scale height $z_d \simeq 350$ pc (see Juri{\'c} et al. 2008 and references therein), 
a fairly cold stellar velocity dispersion, $\sigma \simeq 40$ km s$^{-1}$, and contains stars
that are as old as 10 Gyr (Nordstr{\"o}m et al. 2004).   It remains to be determined whether the Milky Way's
thin disk is typical of spiral galaxies.  This is a vital question.  Unfortunately, scale height measurements for a statistical
sample of external galaxies remain hindered by the presence of absorbing dust lanes in the disk
plane for $\sim L_*$ galaxies (e.g. Yoachim \& Dalcanton 2006).  

Given the uncertainties associated with the formation of disk galaxies, we might make progress by asking 
a few focused, conservative questions.  First, what is the predicted mass range and frequency of large mergers in galaxy-size halos?  
  Second, can a thin stellar disk survive common mergers, and if not, what does this teach us about disk galaxy formation and/or cosmology? 
The figures presented below are taken from work described by Stewart et al. (2008a) on halo merger histories, 
Purcell, Kazantzidis, and Bullock (2008) on stellar disk destruction, and from 
Stewart et al. (2009, in preparation) on the expected gas fractions of mergers.  The simulations 
in Purcell et al. (2008) were motivated by a program developed in Kazantzidis et al. (2008), which aims to
understand the morphological response of disk galaxies to cosmologically-motivated accretion histories.

\begin{figure}[htp]
  \begin{center}
 \includegraphics[width=.8\textwidth]{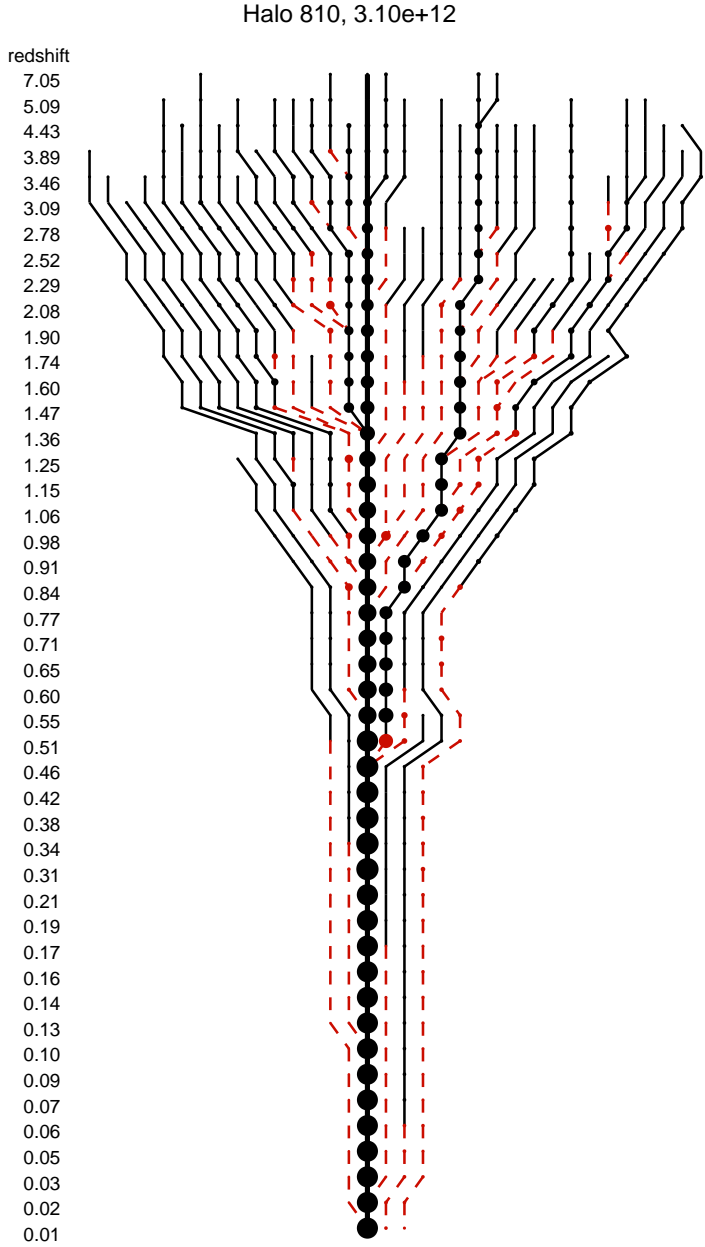} 
  \caption{Sample merger tree for a dark matter halo with $z=0$ mass
   $M_{0} \simeq 3 \times 10^{12} \hMsun$ from Stewart et al. (2008a).  
    Time progresses downward,  with the redshift $z$
  printed on the left hand side.  
 The  bold, vertical line at the center corresponds
 to   the main progenitor,  with  filled  circles proportional to the
  radius of  each halo.  The minimum  mass halo shown in  this diagram
  has   $m =   10^{9.9}  \hMsun$.
 Solid (black) and    dashed (red) lines and    circles
  correspond   to  isolated field  halos,  or  subhalos, respectively.
The dashed (red) lines that do not merge with main progenitor 
represent surviving subhalos at $z=0$.
Note that the halo shown here has a fairly typical merger history, and experiences a merger
of mass $m \simeq 0.1 M_0  \simeq 0.5 M_z$ at $z = 0.51$.}  
\label{mergertrees} \end{center}
\end{figure}

\section{Merger Histories from Cosmological Simulations}

As described in Stewart et al. (2008a), our merger trees are derived from an 80 $h^{-1}$ Mpc box
$\Lambda$CDM simulation.  We concentrate specifically on
thousands of Milky Way-sized systems, $M_0 \simeq 10^{12} \Msun$ at $z=0$.  We categorize
the accretion of objects as small as $m \simeq 10^{10} \hMsun$ and
focus on the infall statistics into main progenitors of $z=0$
halos as a function of lookback time.  

Figure 1 shows a merger tree
for a halo of mass $M_0 = 10^{12.5}
\hMsun$ at $z=0$.  Time runs from top to bottom and the corresponding redshift
for each timestep is shown to the left of each tree.  The radii of the
circles are proportional to the halo radius $R \sim M^{1/3}$, while
the lines show the descendent--progenitor relationship.  The color and
type of the connecting lines indicate whether the progenitor halo is a
field halo (solid black) or a subhalo (dashed red).  The most massive
progenitor at each timestep --- the main progenitor --- is plotted in
bold down the middle.    Once a halo falls within the
radius of another halo, it becomes a subhalo and its line-type changes
from black solid to red dashed.   Figure 1 shows a fairly typical
merger history, with a merger of mass $m \simeq 0.1 M_0$ at $z \simeq
0.51$.  The merger ratio at the time of the merger was $m/M_z \simeq
0.5$.    Note that this large merger does not 
survive for long as  a resolved subhalo --- it quickly loses most of its mass via interactions
with the center of the halo, which presumably would host a central galaxy.  

\begin{figure}[htp]
\begin{center}
\includegraphics[width=0.8\textwidth]{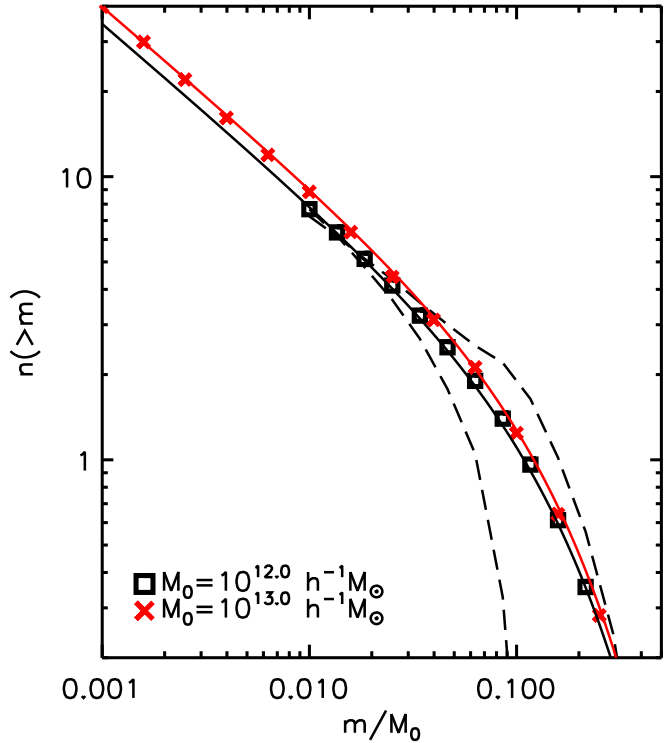} 
\caption{Cumulative mass function of accreted halos from Stewart et al. (2008a).   
The masses of accreted objects have been normalized by the host halo
mass at $z=0$ and the cumulative count is integrated over the
main progenitor's formation history.  
The (black) squares show the average for $10^{12} \hMsun$ halos; (red)
crosses show the average for $10^{13} \hMsun$ halos.  Lines through the
data points show analytic fits provided in Stewart et al. (2008a).  The upper/lower dashed
lines indicate the $\sim 25 \% / 20 \%$ of halos in the $10^{12} \hMsun$
sample that have experienced exactly two/zero $m \geq 0.1 M_0$ merger
events.  Approximately $45 \%$ of halos have exactly one $m \geq 0.1
M_0$ merger event; these systems have mass accretion functions that
resemble very closely the average.}
\label{NgtM}
\end{center}
\end{figure}

Among the most basic questions concerns the mass spectrum of accreted objects.
The solid line in Figure 2
shows the average cumulative number of objects of mass greater than $m$ accreted over 
a halo's history.  Two halo mass bins are shown.   We see that, on average, 
the total mass spectrum of accreted objects (integrated over time) is approximately
self-similar in $z=0$ host mass $M_0$.  
 Milky Way-sized halos
with $M_0 \simeq 10^{12} \Msun$ typically experience $\sim 1$ merger with
objects larger than $m = 0.1 M_0  \simeq 10^{11} \Msun$, and approximately $7$ mergers with
objects larger than $m = 0.01 M_0 \simeq 10^{10} \Msun$ over their histories.
Mergers involving objects larger than $m = 0.2 M_0 \simeq 2 \times 10^{11} \Msun$ should
be extremely rare.

 Figure 3 shows a particularly important statistical summary for the question of
 morphological fractions.  Specifically we show the fraction of galaxy-sized 
 halos (a bin centered on $M_0 = 10^{12} \hMsun$) that have experienced {\em
  at least one} ``large'' merger within the last $t$ Gyr.  The
different line types correspond to different absolute mass cuts on the
accreted halo, from $m > 0.05 \, M_0$ to $m > 0.4 \, M_0$.  
The lines flatten at high $z$ because the halo main
progenitor masses, $M_z$, become smaller than the mass threshold on
$m$.  We find that while fewer than $\sim 10 \%$ of Milky
Way-sized halos have {\em ever} experienced a merger with an object
large enough to host a sizeable disk galaxy, ($m > 0.4 \, M_0 \simeq 4
\times 10^{11} \Msun$), an overwhelming majority ($\sim 95 \%$) have
accreted an object more massive than the Milky Way's disk ($m > 0.05
M_0 \simeq 5 \times 10^{10} \Msun$).  Approximately $70 \%$ of halos
have accreted an object larger than $m/M_0 = 0.1$ in the last 
$10$ Gyr.  We emphasize that the ratios presented here are relative to the
{\em final} halo mass ($m/M_0$) not the ratio of the masses just before
the merger occurred ($m/M_z$).  As presented, the ratios are quite
conservative because halos grow with time
$M_z < M_0$ and $m/M_z > m/M_0$.  We find that typically, for
the mergers we record here, $m/M_z \simeq 2 m/M_0$ (Stewart et al. 2008a)
and that makes the implications for disk survival all the more worrying.

\begin{figure}[t!]
\begin{center}
\includegraphics[width=.95\textwidth]{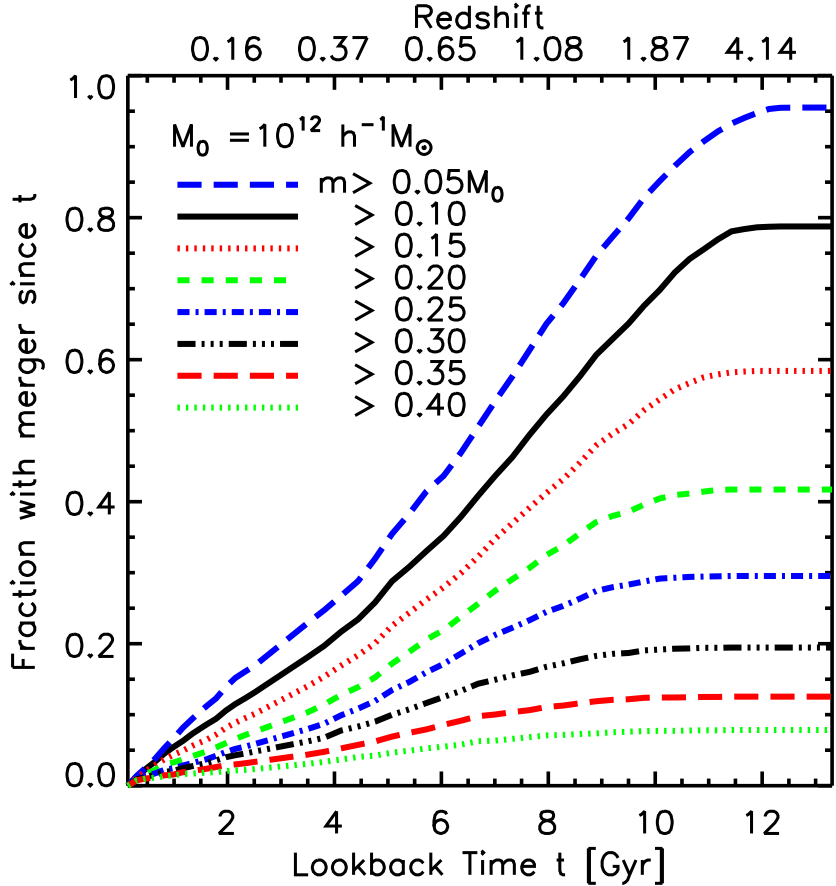}
 \caption{Merger fractions from Stewart et al. (2008a).
 The lines show the fraction of galaxy-sized halos, $M_0 = 10^{12} \hMsun$, 
 that have experienced at least one merger larger than a given mass threshold, $m/M_0$, since look-back time t.
  }
\label{MMfrac} 
\end{center}
\end{figure}

\section{Targeted Simulations}
Recently, Kazantzidis et al. (2008) have investigated the response of galactic disks subject to a $\Lambda$CDM-motivated 
satellite accretion histories and showed that the thin disk component survives, though it is strongly perturbed by the 
violent gravitational encounters with halo substructure (see also Kazantzidis et al., this proceeding). However, these authors focused on subhalos with masses 
in the range $0.01 M_0 \lesssim m \lesssim 0.05 M_{0}$, ignoring the most massive accretion events 
expected over a galaxy's lifetime. Here, we report on the results of Purcell et al. (2008) 
that expand upon this initiative by investigating the evolution 
of galactic disk morphology and kinematics during merger events with mass ratio $m/M_{0} = 0.1$.

As described in more detail in  Purcell et al. (2008), our simulations are performed using the parallel-tree dissipationless code PKDGRAV (Stadel 2001).
 The host halo, disk, and infalling satellites were simulated with $4\times 10^6$, $10^6$ and $10^6$ particles, respectively.  
The primary Milky-Way-analogue system drawn from the set of self-consistent equilibrium models that 
best fit Galactic observational parameters as produced by Widrow et al. (2008), with a host halo mass
of $M_0 = 10^{12} \Msun$ and a disk mass $M_{\rm disk} = 3.6 \times 10^{10} \Msun$.
We initialize a satellite galaxy with a stellar mass of $2 \times 10^9 \Msun$ embedded within a 
dark matter halo of virial mass $m \simeq 0.1 M_{\rm host} = 10^{11} \Msun$.
 In the left panel of Figure 4, we show the edge-on surface brightness map for both 
primary galaxy models, one with a central bulge and one without.

\begin{figure}[t!]
\begin{center}
\includegraphics[width=0.48\textwidth]{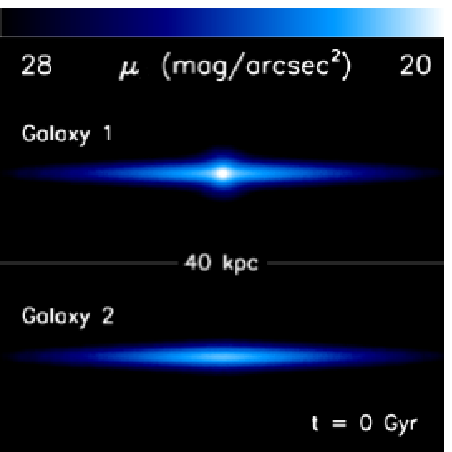}
\includegraphics[width=0.48\textwidth]{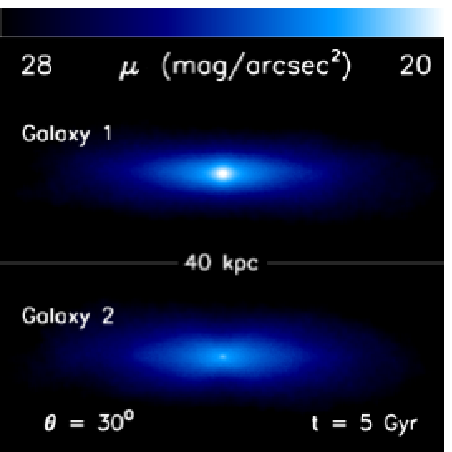} 
\caption{Edge-on surface brightness maps for primary galaxies 1 (upper panels; disk/bulge) and 2 
(lower panels; disk only).  Initial models ($t=0$ Gyr) are shown in the left panel, while the final results 
($t=5$ Gyr) for satellite-infall orbital inclination of $\theta = 30^{\circ}$ appears
on the right panel.  The associated simulations are discussed in Purcell et al. (2008).
}
\label{fig:maps}
\end{center}    
\end{figure}

We explore a range of initial orbital parameters assigned to the merging satellite galaxy, 
motivated by cosmological investigations of substructure mergers (Khochfar \& Burkert 2006; Benson 2005).
We choose an array of orbital inclination angles 
($\theta = 0^{\circ}, 30^{\circ}, 60^{\circ}, \mathrm{and}~90^{\circ}$) in order to assess 
the consequence of this parameter on the evolution and final state of the galactic disk in each case.  
All simulations are allowed to evolve for a total of 5 Gyr, after which time the 
subhalo has fully coalesced into the center of the host halo and the stellar disk has 
relaxed into stability, although there are certainly remnant features in the outer disk and halo that 
will continue to phase-mix and virialize on a much longer timescale; however, our investigations indicate 
that the disk-heating process has reached a quasi-steady state by this point in the merger's evolution.
The morphological thickening of the initial disk after one typical merger is shown in right panels 
of Figure 4.

\begin{figure}[t!]
\begin{center}
\includegraphics[clip,width=0.45\textwidth]{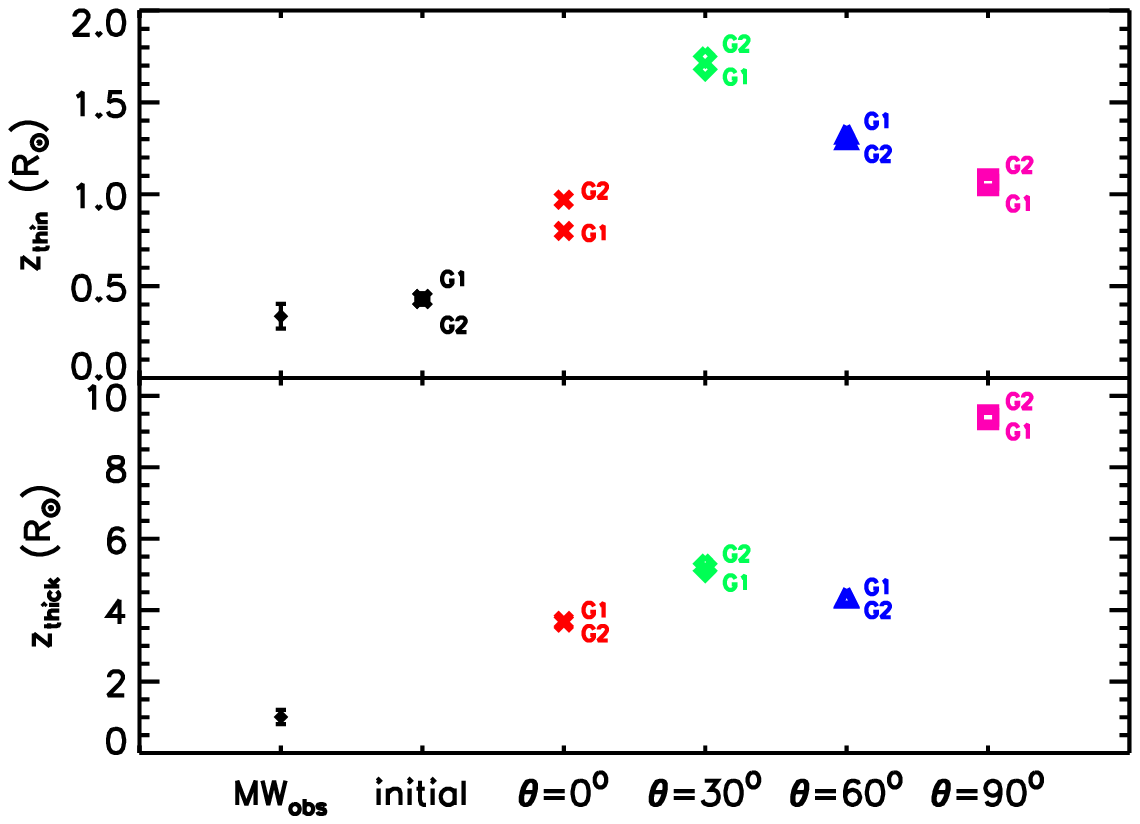}
\hspace{0.3in}
\includegraphics[width=0.4\textwidth]{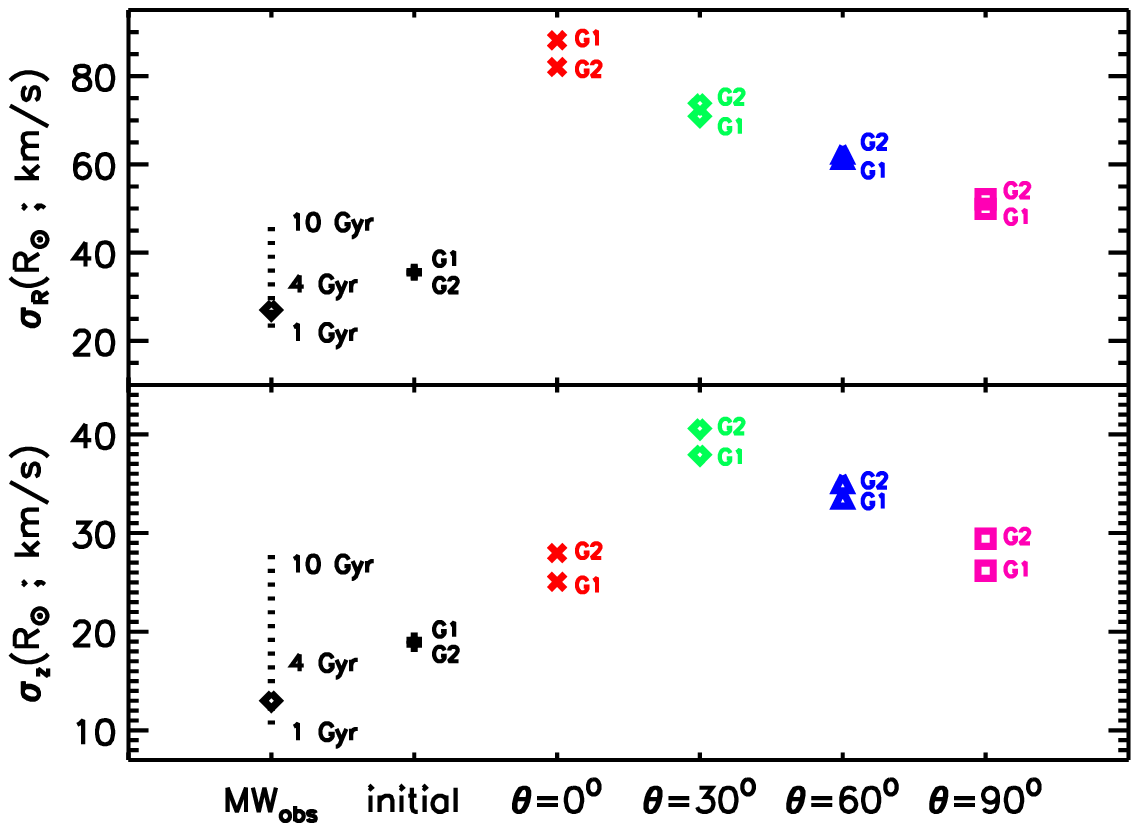}
\vspace{0.1in}
\caption{ {\em Left.} The thin- and thick-disk scale heights in the final state  for each of our simulated 
galaxies from Purcell et al. (2008), compared to the values derived  for the Milky Way. The two panels 
show the result of a two-component sech$^2$ fit, with the upper (lower) panel describing the thin (thick) 
disk's scale height. {\em Right.}  The radial and vertical components of velocity dispersion $\sigma_R$ and $\sigma_z$
at the solar neighborhood ($R=8$ kpc) 
of our simulated disks, compared to the local values obtained by the Geneva-Copenhagen survey.  
In each coordinate, the observational spread is marked by a {\em dotted line} and 
the dispersion of the sample's median-age stars ($t \sim 2-3$ Gyr) is denoted by a {\em diamond}. }
\label{fig:scaleheight}
\end{center}
\end{figure}

Figure 5 provides a direct comparison of all of our merger remnants to observed properties of the Milky Way.
The left panel shows the remnant disk scale heights (derived using two-component fits for a thin and thick disk)
alongside the values  obtained by Juric et al. (2008).  While the thin-disk scale height ($z_{thin}$) of our initial model agrees well with the Galactic 
benchmark of $z_{thin} \simeq 0.3$ kpc, the final systems all have thin-disk components with $z_{thin}$ larger by a factor 
of $\sim 3-5$.  The right panel shows remnant disk velocity dispersions (radial and vertical) as measured in disk planes
around $R = 8$ kpc  compared to the velocity ellipsoid 
observed in the solar neighborhood (Nordstr{\"o}m et al. 2004). 
 Following the simulated 1:10 merger, all three components of velocity dispersion 
are substantially enhanced.  None of the remnants are as cold as the Milky Way disk.  Note that while the 
$\theta = 0^\circ$ in-plane accretion produces the least vertical thickening, it produces a huge amount of radial
heating, and leaves the remnant disk much hotter than that of the Milky Way.  
Our conclusion that cosmologically-motivated 1:10 mergers destroy thin stellar disks.

\begin{figure}[t!]
\begin{center}
  \includegraphics[width=0.95\textwidth]{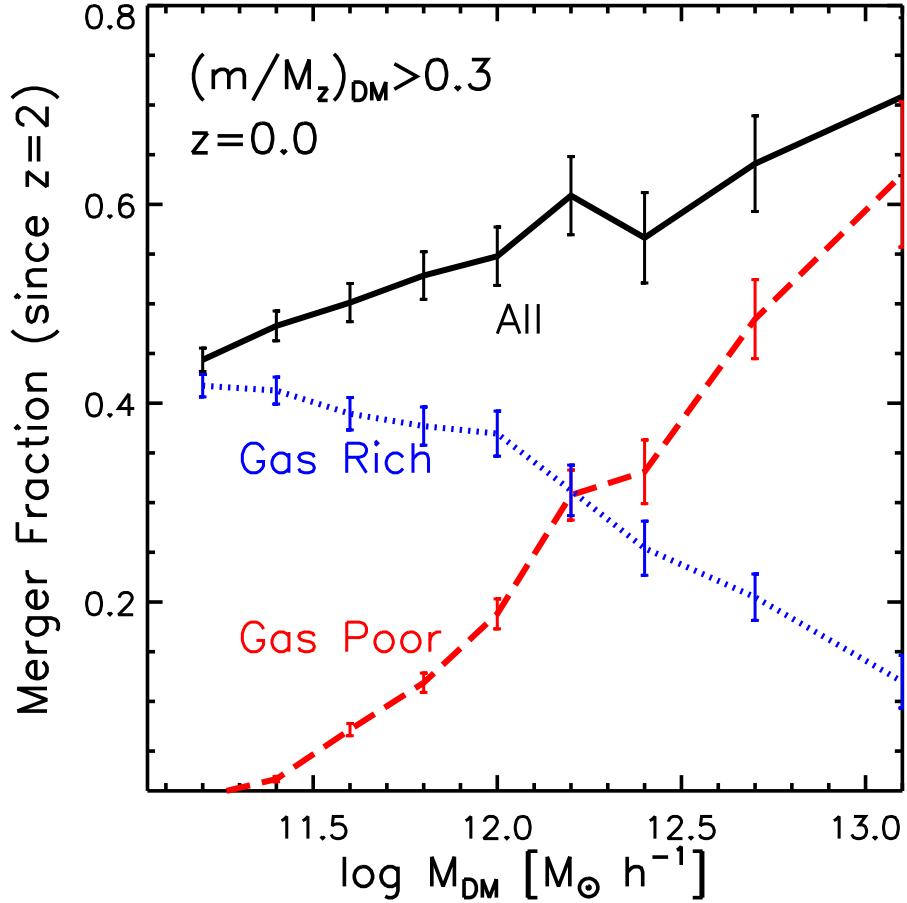} 
    \caption{Fraction of dark matter halos of a given mass that have experienced 
at least one major merger ($m/M_z > 0.3$) since $z=2$.  The solid (black) line is the total 
DM merger fraction, while the dotted (blue) line \emph{only} includes gas rich major 
mergers (see text for details), and the dashed (red) line only includes gas poor major 
mergers.  Error bars are Poisson based on number of host halos and the 
total number of mergers.  The figure is modified from Stewart et al. (2009, in preparation).}
\label{fracz2} 
\end{center}
\end{figure}

\section{Gas-rich Mergers}

As discussed in the introduction, the presence of a stabilizing gas component in merger progenitors can potentially
 alleviate the thin disk disruption we have described.   In order to address  whether it is plausible that gas-rich mergers occur frequently enough to alleviate the problem,
we employ a semi-empirical approach.  Specifically, we assign
stellar masses and gas masses to halos at each of our merger tree timesteps using empirical relations
and then explore the baryonic content of the mergers that occur.
For stellar masses, we use the empirical mapping between halo mass and stellar mass advocated by
Conryoy \& Wechsler (2008).   For gas masses
we use observational relations between stellar mass and gas mass (Kannappan 2004; McGaugh 2005; Erb et al. 2006).
The important qualitative trend is that small halos
tend to host galaxies with high gas fractions and that gas fractions are inferred to increase in galaxies of a fixed
stellar mass at high redshift.

Figure 6 presents an intriguing result from this exploration.
The solid black line shows the fraction of halos that have had a merger larger than
$m/M_z = 0.3$ since $z=2$ as a function of the $z=0$ host mass.  (Note that
here the merger ratio is the ratio of masses just prior to the merger).  
We see that a fairly high fraction ($\sim 60 \%$) of Milky-Way size halos have
experienced a major merger in the last $\sim 10$ Gyr.  
Consider, however, the (blue) dotted  
line, which restricts the merger count to galaxies where {\em both} of the progenitors are
\emph{gas rich} with $f_{\rm gas} = M_{gas}/(M_{gas} + M_*) > 0.5$.
We see that the vast majority of the most worrying mergers in Milky-Way size halos
should have been very gas rich, and that gas-rich major merger become more common
in smaller halos.     The (red) dashed line shows 
the fraction mergers that are made up of more gas poor progenitors with $f_{\rm gas} <  0.5$.
Not only do these trends provide a possible solution to disk survivability, but they 
may also provide an interesting clue to the origin of the
mass--morphology relation.  While dark matter halo merger histories alone show a very weak
trend with halo mass, the baryonic content of the mergers should vary significantly with
halo mass (and by extension, galaxy luminosity).  More massive halos are more likely to experience large, gas-poor mergers, and 
we expect this to result in a higher fraction of spheroid-dominated galaxies.

\bigskip

\noindent {\bf Acknowledgements} \\
\noindent  We thank the conference organizers for an extremely interesting and thought provoking program
and we thank our collaborators for allowing us to present our joint work here.

\end{document}